\documentclass[10pt,twocolumn,preprint]{emulateapj}
\input psfig.sty
\usepackage[english]{babel}
\usepackage{amsmath}

\shortauthors{Zhang et al.} \shorttitle{Increasing the Fisher
Information by Wavelet}

\begin{document}
\title{Increasing the Fisher information content in the matter power spectrum by non-linear wavelet Wiener filtering}

\author{Tong-Jie Zhang$^{1,2,3}$, Hao-Ran Yu$^{1,3}$, Joachim
Harnois-D\'{e}raps$^{3,4}$, Ilana MacDonald$^{3,5}$ and Ue-Li
Pen$^{3}$}

\affil{
$^1$Department of Astronomy, Beijing Normal University, Beijing, 100875, P.~R.~China\\
$^2$Center for High Energy Physics, Peking University, Beijing, 100871, P.~R.~China\\
$^3$Canadian Institute for Theoretical Astrophysics, University of Toronto, M5S 3H8, Ontario, Canada\\
$^4$Department of Physics, University of Toronto, M5S 1A7, Ontario, Canada\\
$^5$Department of Astronomy and Astrophysics, University of Toronto,
M5S 3H8, Ontario, Canada\\}

\email{tjzhang@bnu.edu.cn}

\begin{abstract}
We develop a purely mathematical tool to recover some of the
information lost in the non-linear collapse of large-scale
structure. From a set of 141 simulations of dark matter density
fields, we construct a non-linear Wiener filter in order to separate
Gaussian and non-Gaussian structure in wavelet space. We find that
the non-Gaussian power is dominant at smaller scales, as expected
from the theory of structure formation, while the Gaussian
counterpart is  damped by an order of magnitude on small scales. We
find that it is possible to increase the Fisher information by a
factor of three before reaching the translinear plateau, an effect
comparable to other techniques like the linear reconstruction of the
density field.
\end{abstract}


\maketitle

\section{Introduction}
In modern cosmology, the power spectrum of matter fluctuations,
$P(k)$, is a key measurement, as it is related to many cosmological
parameters. Recent observations such as the cosmic microwave
background \citep{2009ApJS..180..330K}, the correlation function of
galaxies
\citep{2005ApJ...619..178E,2006PhRvD..74l3507T,2007ApJ...657...51P,2007MNRAS.381.1053P,2010MNRAS.401.2148P},
the strong gravitational lensing probability for image separation
\citep{2003ApJ...587L..55C,2006ApJ...650L...9C,2006MNRAS.367.1241O},
and the shear correlation function for weak gravitational lensing
\citep{2003MNRAS.346..994P,0004-637X-647-1-116,2007MNRAS.381..702B,2008A&A...479....9F},
have been able to but constraints on many parameters. In many
analyses, cosmologists are interested in measuring the matter power
spectrum on large scales, which holds a wealth of cosmological
information that has not been erased by the non-linear gravitational
collapse. The constraining strength of a survey depends directly on
the amount of Fisher information
\citep{Fisher1935,1997ApJ...480...22T}, i.e. statistically
independent Fourier modes, contained in the measurement, and one
needs to maximize the information in order to minimize the
uncertainty on cosmological parameters.

\cite{2005MNRAS.360L..82R,2006MNRAS.371.1205R} first studied the
amount of Fisher information as a function of scale contained in the
matter power spectrum. They estimated the information content about
the initial amplitude of the linear power spectrum from an ensemble
of many $N$-body simulations and showed that the Fisher information
is preserved in the matter power spectrum on large, linear scales.
They also observed that the information plateaus in the translinear
regime ($k \sim 0.2-0.8 h/{\rm Mpc}$), and that there is a second
increase in information when including the small, non-linear scales.
\cite{2006MNRAS.370L..66N} then found that the halo model also
predicts this behavior on translinear scales in the information
content, which comes largely from cosmic variance in the number of
the largest haloes in a given volume. Furthermore,
\cite{2007MNRAS.375L..51N} found that the cosmological Fisher
information in the matter power spectrum about many key cosmological
parameters is quite degenerate both on translinear and non-linear
scales. They also showed that it can be difficult to constrain the
initial cosmological conditions from the dark matter power spectrum
on non-linear scales, and suggested the removal of the largest
haloes as an alternative to resolve this problem. Based on the
technique of \cite{2005MNRAS.360L..82R,2006MNRAS.371.1205R},
\cite{2008ApJ...686L...1L} estimate the Fisher information in the
galaxy angular power spectrum from the photometric redshift
catalogue of the Sloan Digital Sky Survey Data Release 5. Their
results also show that very little new information is contained in
the matter density power spectrum on the translinear scale, which is
observationally consistent with the previous numerical trend
\citep{2005MNRAS.360L..82R,2006MNRAS.371.1205R,
2006MNRAS.371.1188H,2006MNRAS.370L..66N,2007MNRAS.375L..51N}.

Many methods have been proposed to recover parts of this lost
information, notably running $N$-body simulations backwards in time
\citep{2000ASPC..201..282G}, or density field reconstruction from
linear theory
\citep{2007ApJ...664..675E,2009PhRvD..80l3501N,2009PhRvD..79f3523P}.
\cite{2009ApJ...698L..90N} introduces a method of logarithmic
mapping of the density field, successfully recover great amount of
information content in the dark matter power spectrum.
\cite{1992MNRAS.254..315W} uses a method called Gaussianization,
which is a monotonic transformation of the smoothed galaxy
distribution to reconstruct primordial density fluctuations. Here we
attempt to Gaussianize the density fields by non-linear Wiener filtering. Our
objective is to filter out the non-linear structure, leaving behind
a density field in which the amount of Fisher information has
increased.

Structure formation in current theories starts from an initially
linear Gaussian random field, which progressively becomes non-linear
through gravitational instabilities, starting at the smallest
scales. This process is local in real space, and only the linear
part of the field remains intrinsically Gaussian. In the non-linear
regime, the Fourier modes start to couple together, and the field is
no longer Gaussian. It is legitimate to ask how to filter out the
non-linearities. Wavelet functions provide a natural basis that
compromises between the locality in real and Fourier space.
Moreover, it can simultaneously decompose the data, functions or
even operators in coordinate (or time) space and scale (or
frequency) space. Therefore, wavelet transforms are an ideal
numerical technique for extracting multi-scale information
\citep{1998World.Sci..2A}, as required by our Wiener filter.
\cite{1996ApJ...459....1P} applied this method to the high redshift
L$\alpha$ systems, which often have separations on translinear
scales. \cite{1999RSPTA.357.2561P} also discussed the application of
wavelets to filter the Gaussian noise contained in some images, and
found that the wavelet basis is highly advantageous over the
standard Fourier basis if the data are intermittent in nature.

In this paper, we apply the wavelet method in order to push the
translinear plateau further in the information content, without
removing the largest haloes in the analysis. The outline of the
paper is as follows. In \S 2, we present the $N$-body code that was
used to simulate the dark matter density fields. In \S 3, we discuss
the application of the discrete wavelet transform (DWT) and Bayesian
theory to filter out some of the non-Gaussian component of the
density fields. In \S 4, we compare the power spectra of the
filtered and unfiltered data, while in \S 5 we analyze the
information content in both data sets. Discussion and conclusions
are presented in \S 6.

\section{$N$-body Simulations of dark matter density fields}
We use the cosmological simulation code CubeP3M (CITA Computing
2008), an upgraded version of PMFAST \citep{2005NewA...10..393M}
that uses a cubical decomposition instead of slabs in the global
FFT, and thus scales to much larger runs. In addition, the
gravitational force is now calculated down to the subgrid level,
which increases the code resolution. Finally, it is both MPI and
openmp parallelized, perfect for running on most computer clusters.

We run 141 simulations with a box size of 600 $h^{-1}$ Mpc and a
resolution of $512^3$ cells and $256^3$ particles. The
initialization step of the simulation first reads a CMBFAST
\citep{1996ApJ...469..437S} transfer function, which models our
fiducial cosmology, and evolves the power linearly to an initial
redshift of $z=200$. It then uses the Zel'dovich approximation to
calculate a displacement field and a velocity field, which are
assigned to the initial particles. The cosmological parameters used
are $\Omega_{\rm M}=0.29$, $\Omega_\Lambda=0.71$, $h=0.73$,
$\sigma_8=0.84$, and $n_s=0.95$. The code then evolves these initial
densities up to $z=0$. One of those density fields is plotted in
Fig.\ref{fig:maps}(a). For the purpose of visualization, the 3D
field is projected onto a 2D plane by averaging over part of the
third dimension.

\section{Non-Gaussian filtering of density fields}

\subsection{Discrete wavelet transform (DWT)}

Similar to the Fast Fourier Transform (FFT), the Discrete Wavelet
Transform (DWT) is an invertible, linear operation, and can be
considered as a rotation in function space. Both operations
transform a data vector, whose length must be an integer power of 2,
to a numerically different vector of the same length. The FFT's
basis functions are sines and cosines, which, of course, are not
localized in real space. The DWT, on the other hand, has two kinds
of basis functions, called scaling functions (mother functions) and
difference functions (wavelet functions). Unlike sines and cosines,
individual DWT basis functions are localized both in real and
frequency space, and because they vary both in scales and locations,
they enable us to analyze more accurately the fine variations in a
data set.

Although there is an infinite number of possible forms of DWT basis
functions, only a few have a clear meaning. The simplest discrete
wavelet is the Haar wavelet, which suffers from the fact that its
basis functions are discontinuous and provide a poor approximation
to smooth functions. We consider in this paper the Daubechies-4
wavelet basis (hereafter DB-4) \citep{Daubechies1992}, which
contains families of scaling functions and wavelets that are
orthogonal, continuous and have compact support.

We denote the one-dimensional scaling function and wavelet function
as $\phi$ and $\psi$. For simplicity we first consider a
one-dimensional vector $\mathbf{d}(\vec{x})$ with length $L=2^J$.
Using DB-4 wavelet, it can be expanded as
\begin{equation}\label{expansion}
    d(x)=\sum_{l=0}^{1}\epsilon_{0,l}\phi_{0,l}(x)
    +\sum_{j=0}^{J-1}\sum_{l=0}^{2^j-1}
    \tilde{\epsilon}_{j,l}\psi_{j,l}(x),
\end{equation}
where $j$ and $l$ can be regarded as a dilation and a translation of
the basis function:
\begin{equation}\label{}
    \phi_{j,l}(x)=\sqrt{\frac{2^j}{L}} \phi(2^jx/L-l)
\end{equation}
\begin{equation}
    \psi_{j,l}(x)=\sqrt{\frac{2^j}{L}} \psi(2^jx/L-l).
\end{equation}
$\epsilon_{j,l}$ and $\tilde{\epsilon}_{j,l}$ are called scaling
function coefficients (hereafter SFCs) and wavelet function
coefficients (hereafter WFCs). They can be calculated by
\begin{equation}\label{}
    \epsilon_{j,l}=\int d(x)\phi_{j,l}(x){\rm d}x,
\end{equation}
and
\begin{equation}
    \tilde{\epsilon}_{j,l}=\int d(x)\psi_{j,l}(x){\rm d}x.
\end{equation}
Any particular set of wavelets is specified by a corresponding set
of wavelet filter coefficients. The most localized Daubechies
wavelet is DB-4 wavelet. Its scaling and wavelet functions have 4
coefficients,  $a_0,...,a_3$ and $b_{-1},...,b_1$ respectively.
\begin{eqnarray*}
  a_0=(1+\sqrt3)/4, \ \ \ \ a_1=(3+\sqrt3)/4, \\
  a_2=(3-\sqrt3)/4, \ \ \ \ a_3=(1-\sqrt3)/4,
\end{eqnarray*}
and
\begin{eqnarray*}
  b_{-2}=(1-\sqrt3)/4, \ \ \ \ b_{-1}=(\sqrt3-3)/4, \\
  b_0=(3+\sqrt3)/4, \ \ \ \ b_1=(\sqrt3-1)/4.
\end{eqnarray*}
The wavelet transform of a vector of data is done first by applying
a wavelet coefficient matrix to the whole vector, and then applying
a ``smaller" matrix to the smoothed vector of length $L/2$. For DB-4
wavelet transform, this was done until only two SFCs were left.
Finally, two SFCs store the information of the mean of the largest
scale of the vector, and $L-2$ WFCs store the information of the
fluctuations or the differences between regions of the vector. For
3D simulation data, the process is similar. The data is wavelet
transformed in one arbitrary direction $x_1$, the result is
transformed in another direction $x_2$, and then in $x_3$. Finally,
the result contains $2^3=8$ SFCs and $512^3-8$ WFCs.

\subsection{Non-linear filtering}

We now have the DWT transform of those density fields which we need
to filter. Because of the linearity of wavelet transforms, we can
think of the data as being filtered into two pieces, the inverse
wavelet transforms of which are called the Gaussianized density
$\mathbf{d}_a$ and the non-Gaussianized density $\mathbf{d}_b$. The
unfiltered density fields $\mathbf{d}$ consist of the sum of these
two parts, and the hope is that the Gaussianized density will be
closer to linear theory. Since the filtering process is in wavelet
space, we divided the wavelet transform of the density field
$\mathbf{D}$ into the Gaussianized part $\mathbf{A}$ and the
non-Gaussianized part $\mathbf{B}$. Again, because of the linearity
of wavelet transforms, we have
\begin{equation}\label{dab}
    \mathbf{d}=\mathbf{d_{a}}+\mathbf{d_{b}}
\end{equation}
in real-space and
\begin{equation}\label{dabw}
    \mathbf{D}=\mathbf{A}+\mathbf{B}.
\end{equation}
in wavelet space.

The method of thresholding and Gaussian Wiener filtering is not
suitable in this work, because we need to filter localized
structure, so we use a non-Gaussian Wiener filter instead. In
wavelet basis the non-Gaussianity is clearly represented
\citep{1999RSPTA.357.2561P}, so the filter is applied on each
wavelet modes. Each wavelet mode has three $j$ 's to characterize
the mode's scale, and we denote wavelet modes with $\mathbf{s}$. We
regard $\mathbf{A}$ as Gaussian white noise $\mathbf{N}$, as in
\cite{1999RSPTA.357.2561P}, $\mathbf{B}$ as the original image
$\mathbf{U}$, and $\mathbf{D}$ as the noisy image
$\mathbf{D}=\mathbf{A}+\mathbf{B}$. For each wavelet mode, we assume
that the WFCs $A(\mathbf{r},\mathbf{s})$, at fixed sclae $\mathbf{s}$,
has the form of a Gaussian distribution $\mathcal{N}(0,\sigma^2)$ over
position $\mathbf{r}$, and if we know the PDF of
$D(\mathbf{r},\mathbf{s})$, all the modes are statistically
independent. We denote the PDF for a given mode of
$B(\mathbf{r},\mathbf{s})$ as $\Theta(b)$, and we obtain the
variable $D(\mathbf{r},\mathbf{s})$ with PDF
\begin{equation}\label{fd}
    f(d)=\frac{1}{\sqrt{2\pi}}\int\Theta(b)\exp(-\frac{1}{2}(d-b)^2){\rm d}b
\end{equation}
According to Bayes's theorem, we can calculate the conditional
probability $P(B|D)=P(D|B)P(B)/P(D)$. For the posterior conditional
expectation value, we get
\begin{equation}\label{s}
    \langle B|D=d \rangle=\frac{1}{\sqrt{2\pi}f(d)}\int{\exp[-\frac{1}{2}(u-d)^2]\Theta(u)u{\rm d}u}
\end{equation}

\begin{equation}\label{beyes}
    B=D+\frac{1}{\sqrt{2\pi}f(d)}\partial_d\int{\exp[-\frac{1}{2}(u-d)^2]\Theta(u){\rm d}u}
\end{equation}
\begin{equation}\label{beyes}
    B=D+(\ln f)'(d).
\end{equation}
We define two filter functions
\begin{equation}\label{dalpha}
    \alpha(\mathbf{s})=A/D,
\end{equation}
\begin{equation}\label{dbeta}
    \beta(\mathbf{s})=B/D,
\end{equation}
so
\begin{equation}\label{alpha}
    \alpha(\mathbf{s})=-\frac{(\ln f)'(d)}{d},
\end{equation}
\begin{equation}\label{beta}
    \beta(\mathbf{s})=1+\frac{(\ln f)'(d)}{d}.
\end{equation}
In order to realize the filtering of the actual data, we have to
estimate the prior distribution $\Theta$ for each wavelet mode.

We assume that the one-point distribution of modes is non-Gaussian
and that the modes are statistically independent, and so we need to
specify the PDF for each mode. In this work we use Cartesian product
wavelets \citep{Meyer1992}. This basis for our density fields has
three scales $2^{J-j_1}$, $2^{J-j_2}$ and $2^{J-j_3}$ (in unit of
pixels), which are the length, width and height of the cuboid each
wavelet support in real-space. We define the volume of the cuboid in
real-space as the equivalent volume $V_e$. For the real data, the
structures dealt with have different physical scales and specific
locations in the 3D density field. Consequently, the optimal basis
must also depend both on the scale and the location. As mentioned
before, our wavelet basis, which is localized both in frequency
space and in real-space, are perfectly well suited to effectively
extracting the information of these structures. Every WFC can be
seen as the dot product of the density field and a wavelet of a
particular scale and shape.

Since the modes are statistically independent, it is natural to collect
together all the WFCs with the same corresponding cuboids, and
then to obtain the PDF of this scale. Actually, we do not differentiate
the direction of the cuboids and consider only the combination of
$j_1$, $j_2$ and $j_3$. That is to say, we regard
$(j_1,j_2,j_3)=(1,2,3)$ and $(j_1,j_2,j_3)=(1,3,2)$ as the same cuboids,
and the corresponding WFCs are collected together. Note that there
may be more than one combination whose equivalent volumes are the
same.

\begin{figure}
    \epsscale{1.3}
    \plotone{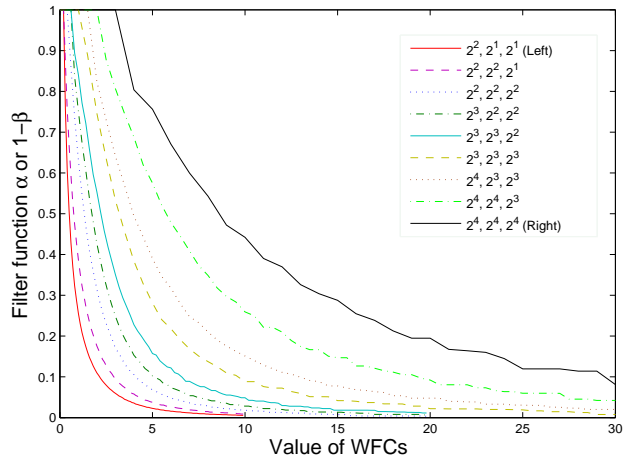}
    \caption{Filter functions $\alpha$ and $\beta$ as a function of the
    WFC's value, 9 wavelet modes are chosen. Each mode is described
    by a combination of three lengths, as seen in the legend, which form the
    cuboid that the DWT basis function governs. Moving from left to right in
    the plot corresponds to moving from top to bottom in the legend. As WFCs
    are small, one can see that $\alpha$ is simply unity, where Gaussianity dominates.}
    \label{fig:fcurve}
\end{figure}

\begin{figure}
    \epsscale{1.3}
    \plotone{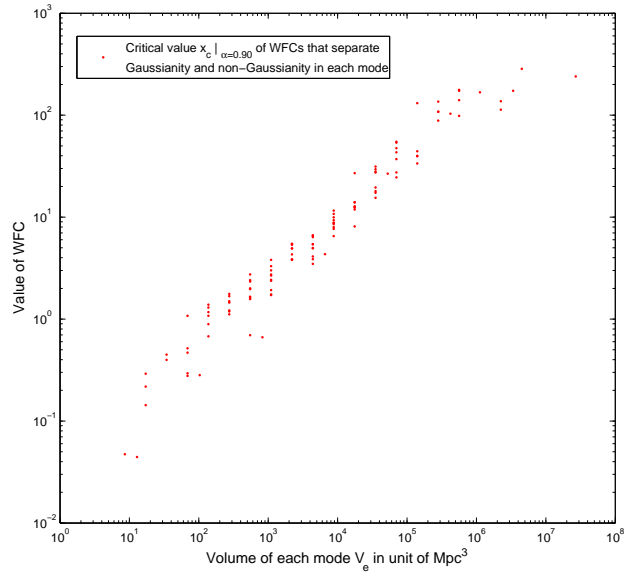}
    \caption{Critical value of WFC $x_c$ of the scales versus the equivalent
    volume $V_e$ of the basis function. The $x_c$ is defined via
    $\alpha(x_c)=0.9$, which can be regard as a
    boundary between Gaussianity and non-Gaussianity on that scale.}
    \label{fig:xthres}
\end{figure}

However, fewer WFCs are available for larger scales, or greater
equivalent volumes, and thus the PDF has lower statistics, which
makes the calculation of filter functions less accurate. If there
are not enough WFCs, or if improper statistics are used, the
appearance of fluctuation in the PDF, especially the region near
zero where the property of Gaussianity is detected, reduces our
precision in the determination of the filter functions. However,
since we collect the WFCs over 141 density fields we have enough
statistics to make both the PDF and filter functions smooth. Also,
in counting the WFCs, we choose an adequate set of intervals, or
bins, in order to best represent the shape of the PDF. The range of
the intervals must both be large enough to contain the non-Gaussian
`tails' in the PDF, and small enough to show the detailed properties
of Gaussianity. We discuss these in the next paragraph. We find that
the WFCs on different scales have different distributions. The
curves of PDFs where WFC's value are relatively small appear to be
Gaussian, but there are non-Gaussian `tails' for larger values of
PDFs. It seems that if those WFCs belong to larger scales, or larger
$V_e$, the `boundary' value between Gaussianity and non-Gaussianity
is larger. We plot 9 filter functions in Fig.\ref{fig:fcurve} to
compare the shape of the filter function curve between scales.  This
property is more easily visualized if we define a critical value on
each scale, defined as the critical value of WFC $x_c$ that yields $\alpha=0.9$. In
Fig.\ref{fig:xthres} we plot that critical value $x_c$ versus $V_e$.

The CDF of a set of data is obtained by sorting and by doing
derivatives to get the PDF. However, it is challenging to sort all
the WFCs on one scale among hundreds of density fields. Instead, we
first construct several hundreds of interval bins, and then read
every density field datum to check the WFCs on each scale, and find
which interval each of these WFCs belongs to. Actually, there are a
total 120 different scales, so we need to count all the WFCs on each
scale and on each of the 141 densities to produce our final 120 PDFs
of WFCs. Then, we check each of these PDFs, and determine whether
the intervals properly represent the shape of PDFs. If not, we may
try another series of bins, get a new set of PDFs, and iterate until
all the PDFs have been measured.

We also use i) the same set of simulations but with lower
resolution, and ii) new set of simulations with finer resolution,
and get similar PDFs of WFCs (thus filter functions) on same
physical scales. As a result, the resolution of simulations does not
affect WFCs' PDF if the physical scales are the same.

From the relationship between the critical values and $V_e$, we can
see that for smaller scales, the critical values are smaller, such
that the signal contributes more to the non-Gaussian density field,
as expected.

\begin{figure*}
    \centerline{
    \psfig{figure=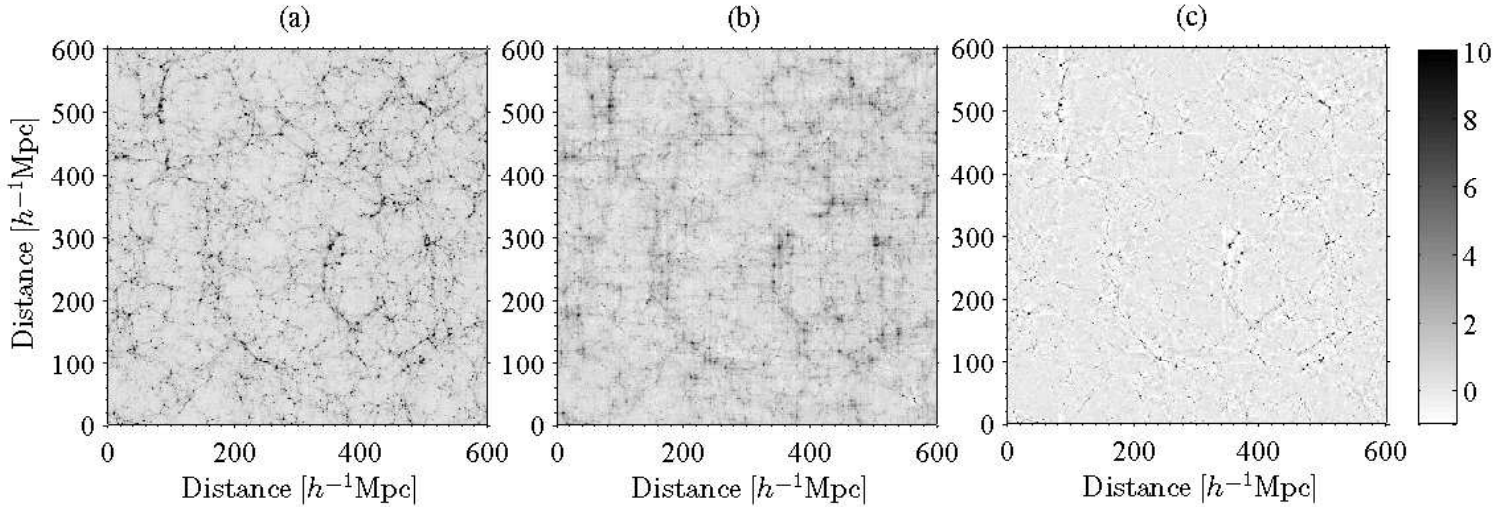,width=8.2truein,height=3.1truein}
    \hskip 0.0in}
    \caption{\textbf{(a)} Map of a density field with a width of 600
    $h^{-1}$ Mpc and $512^3$ pixels, randomly selected from 141 $N$-body
    simulations. A slice of the density field is projected onto the x-y
    plane for visualization. \textbf{(b)} Same as \textbf{(a)}, but the
    map is Gaussianized by wavelet filtering. \textbf{(c)} Same as
    \textbf{(a)}, but the map is non-Gaussianized by wavelet filtering.
    \textbf{(b)} and \textbf{(c)} add up to \textbf{(a)}.}
    \label{fig:maps}
\end{figure*}

We finally filter all the data by multiplying the densities in
wavelet space with our filter functions, Eq.\ref{alpha} and
Eq.\ref{beta}, which separates those 141 density fields into
Gaussian and non-Gaussian contributions. Note that we do not filter
the SFCs because they only govern the mean of the largest scales and
have no significant impact on the power spectra. We plot a slice of
an unfiltered density and its Gaussianized and non-Gaussianized
components in Fig.\ref{fig:maps}.

\section{Comparison of power spectra}
The power spectrum  is the Fourier transform of the correlation
function and measures the amount of clustering in the matter
distribution. It is measured in terms of the wavenumber $k$ in units
of $h {\rm Mpc}^{-1}$. In the case of a  Gaussian random field, the
power spectrum describes  completely the statistics of the
distribution, as all high moments either vanish or factor into its
explicit functions. The power spectrum from each simulation is
defined as
\begin{equation}\label{}
    \langle|\delta(\vec{k},\vec{k'})|\rangle^2=(2\pi)^3P(\vec{k})\delta(\vec{k}-\vec{k'}),
\end{equation}
where $\langle|\delta(\vec{k},\vec{k'})|\rangle^2$ is the volume
average of the expectation value and $P(\vec{k})$ is the power
spectrum. Of equal interest is $\Delta^2_k$,  the power spectrum in its dimensionless
form, defined as:
\begin{equation}\label{}
    \Delta^2_k\equiv\frac{k^3P(k)}{2\pi^2}.
\end{equation}

\begin{figure}
    \epsscale{1.3}
    \plotone{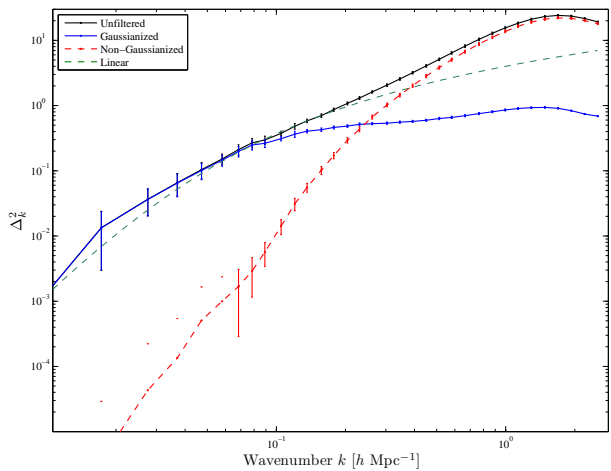}
    \caption{Power spectra and error bars of the density field simulations, and their
    Gaussianized and non-Gaussianized power spectra. Each curve is
    determined by the mean of over 141 power spectrum samples, and the
    error bars describe those standard deviations. The error was
    multiplied by 3. The BBKS linear power spectrum is plotted in dash line.} \label{fig:ps}
\end{figure}

For the 141 simulations, Gaussianized and non-Gaussianized densities
are generated, as described in the previous section. The power
spectra of these 423 mass distributions are calculated using the
`Nearest Grid Point' (NGP) mass assignment scheme, which calculates
the position of each particle based on which grid point it is
nearest. In Fig.\ref{fig:ps} we plot the mean power spectrum and
error bars of 141 unfiltered, Gaussianized and non-Gaussianized
density fields. We can see that for small $k$, corresponding to
large scales, Gaussianized fluctuations dominate the power, and that
for large $k$, corresponding to small scales, non-Gaussianized
fluctuations dominate. Non-Gaussianized fluctuations begin to
dominate the power in the region $k\sim0.2$, which is the boundary
of the linear power spectrum and the non-linear power spectrum. We
can see that the non-Gaussian filter in wavelet space is quite
efficient at separating the linear and non-linear power spectra.

\section{Information content in the power spectra of density fields}

The covariance is the mean value of the product of the deviations of
two variables from their respective means. Simply put, it measures
the correlation between the variance, or the error bars, of the
power spectra at different scales $k$. If the measurements were
completely uncorrelated, the diagonal of the covariance matrix would
be the variance at each value of k, and all the off-diagonal entries
would be zero. Mathematically, the covariance matrix is defined as
\begin{equation}
    C(k,k')\equiv \frac{1}{N-1}\sum_{i=1}^N
    [P_i(k)-\langle P(k)\rangle][P_i(k')-\langle P(k')\rangle],
\end{equation}
where $N$ is the number of realizations and $\langle P(k)\rangle$
is the mean power spectrum.

\begin{figure*}
    \centerline{
    \psfig{figure=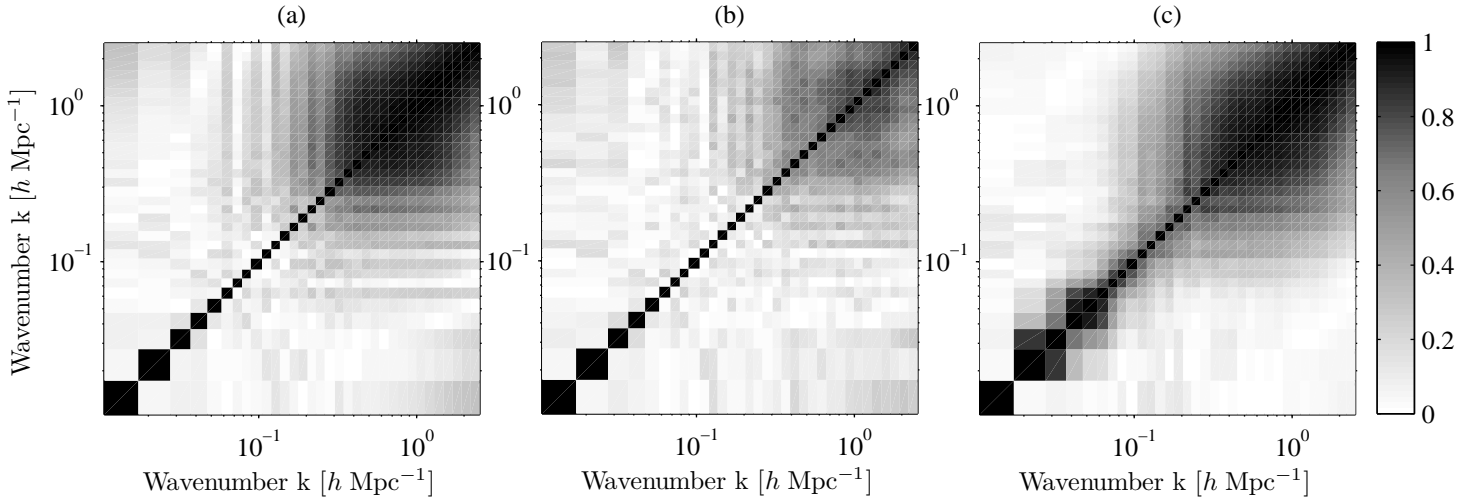,width=9truein,height=3truein}
    \hskip 0.0in}
    \caption{\textbf{(a)} Cross-correlation coefficient
    matrix as found from 141 power spectra of the density field
    simulations. The squares in black on the diagonal line indicate
    perfect correlation. \textbf{(b)} Same as \textbf{(a)}, but the
    power spectra are from the Gaussianized density fields. \textbf{(c)}
    Same as \textbf{(a)}, but the power spectra are from the
    non-Gaussianized density fields.\\
    \\
    \\}
    \label{fig:matrices}
\end{figure*}

The cross-correlation coefficient matrix, or for short the
correlation matrix, is a normalized version of the covariance
matrix, where each value is divided by the square root of the
diagonal values as follows:
\begin{equation}\label{rho}
    \rho(k,k')= \frac{C(k,k')}{\sqrt{C(k,k)C(k',k')}}.
\end{equation}
The three correlation matrices for our unfiltered, Gaussianized and
non-Gaussianized densities are shown in Fig.\ref{fig:matrices}. For
the covariance matrix of the unfiltered densities, the linear regime
is diagonal, while in the non-linear regime, the power spectra at
different $k$-modes are correlated by at least 40\%. In the Gaussian
filtered covariance matrix, however, we find a smaller correlated
region. It seems that the correlations between $k$-modes are
suppressed to some extent. When we look at the non-Gaussian filter,
we see that the diagonal is now a few cells thick, i.e. some
non-diagonal elements are correlated even in the linear regime. This
is allowed, since the `Gaussian + non-Gaussian' decomposition is
done on the densities, not on the covariance matrix. Also, the
corresponding value in the covariance matrix are many orders of
magnitude smaller than those in the unfiltered and Gaussian cases,
and thus play no role what so-ever.

\begin{figure}
    \epsscale{1.2}
    \plotone{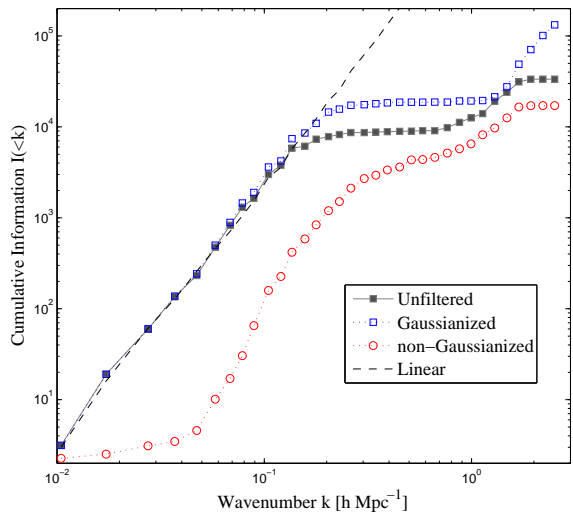}
    \caption{Cumulative
    information in the power spectra as a function of the wavenumber.
    The black squares filled with grey correspond to the density field
    simulations without filtering. The blue squares and the blue solid
    line correspond to the Gaussianized densities. The red squares and
    the red dashed line correspond to the non-Gaussianized densities.}
    \label{fig:info}
\end{figure}

The cumulative Fisher information is defined as follows: for a given
wavenumber, we select a subsection of the covariance matrix to that
wavenumber, and we invert this sub-matrix, and sum over all its
elements. The normalized covariance matrix is described as follows
\begin{equation}\label{Cnorm}
    C_{\rm norm}(k,k')=\frac{C(k,k')}{\langle P(k)\rangle \langle
    P(k')\rangle},
\end{equation}
and our cumulative information function is
\begin{equation}\label{cumuinfo}
    I(k_n)=\sum_{i,j=1}^n C_{\rm norm}^{-1}(k_i,k_j).
\end{equation}
We plot the cumulative information of the unfiltered, Gaussianized
and non-Gaussianized power spectra in Fig.\ref{fig:info}. One can
see that in the translinear regime, where $k\sim0.2$, the cumulative
information of the density fields has a flat plateau, which
indicates that there is nearly no independent information in the
non-linear regime of the power spectrum. In comparison, for the
Gaussianized information, the cumulative information also has a
plateau, but at a higher value, which indicates that it contains
about three times more independent Fourier modes. This is a key
result, especially for parameter estimation obtained through Fisher
matrices. These analyses could be improved by up to a factor of
$\sqrt{3}$ solely by filtering their data.

\section{Conclusion}
We use the code `CubeP3M' to generate 141 dark matter density
fields. In wavelet space, we construct a non-Gaussian Wiener filter
in order to separate non-Gaussian and Gaussian structures. We
measure the power spectra of the unfiltered, Gaussianized and
non-Gaussianized density fields, and find that the Gaussianized
power spectrum is heavily dampened on small scales, while the
non-Gaussian features dominate at smaller scales, as expected from
structure formation. We also calculate the cumulative information,
and find that the plateau in the translinear regime rises in the
Gaussianized information curve by a factor of three.

\acknowledgements

We are very grateful to the anonymous referee for many valuable
comments that greatly improved the paper. We thank Peng-Jie Zhang
and Cong Ma for their friendly helps and suggestions. This work was
supported by the National Science Foundation of China (Grants No.
10473002), the Ministry of Science and Technology National Basic
Science program (project 973) under grant No. 2009CB24901, the
Fundamental Research Funds for the Central Universities. The
simulations were performed on the Sunnyvale cluster at CITA. UP and
JHD would like to acknowledge NSERC for their financial support.

\bibliography{m}
\bibliographystyle{hapj}

\end{document}